\documentclass[longbibliography,titlepage,secnumarabic,amssymb, nobibnotes, aps, pre,reprint]{revtex4-1}
\usepackage{graphicx}
\usepackage{caption}
\usepackage{amsmath,amssymb}
\usepackage[mathlines]{lineno}
\usepackage{hyperref}
\begin{document}

\title{Two jets during the impact of viscous droplets onto a less-viscous liquid pool}%
\author{Quan Ding}
\author{Tianyou Wang}
\author{Zhizhao Che}
\email[]{chezhizhao@tju.edu.cn}
\affiliation{State Key Laboratory of Engines, Tianjin University, Tianjin, 300072, China.}
\date{\today}%
\begin{abstract}
It is generally accepted that the Worthington jet occurs when a droplet impacts onto a liquid pool. However, in this experimental study of the impact of viscous droplets onto a less-viscous liquid pool, we identify another jet besides the Worthington jet, forming a two-jet phenomenon. The two jets, a surface-climbing jet and the Worthington jet, may appear successively during one impact event. By carefully tuning the impact condition, we find that the two-jet phenomenon is jointly controlled by the droplet-pool viscosity ratio, the droplet Weber number, and the droplet-pool miscibility. The mechanism of the surface-climbing jet is completely different from that of the Worthington jet: the liquid in the pool climbs along the surface of the droplet and forms a liquid layer which converges at the droplet apex and produces the surface-climbing jet. This surface-climbing jet has a very high speed, i.e., an order of magnitude higher than the droplet impact speed. The effects of the impact speed, droplet viscosity, droplet size, and surface tension on the surface-climbing jet are also analysed. This study does not only provide physical insights into the mechanism of droplet and jet dynamics, but also be helpful in the optimisation of the droplet impact process in many relevant applications.
\end{abstract}
\maketitle

\section{Introduction}\label{sec:sec1}
The dynamics of droplet impact on surfaces is of great importance in a large variety of industrial applications, and the outcomes of the impact process directly affect the performance of these applications or even result in catastrophic accidents, such as in spray cooling, internal combustion engines, droplet-impact erosion on turbine blades, ink-jet printing, and bloodstain pattern analysis. The droplet impact process has attracted much attention not only because of its practical applications but also due to its scientific significance as a fundamental phenomenon of fluid mechanics. The impact of droplets onto a liquid pool appears in many natural processes. For example, the large-scale and widespread natural phenomenon of the aeration of the surface layers of lakes, seas, and oceans depends on air bubble entrainment due to raindrop impacts \cite{H.N.Oguz1990BubbleEntrainment}. These impacts on ocean surfaces lead to the formation of upward jets and secondary droplets, which evaporate and form salt crystals. The study of droplets impacting onto a liquid pool is also important in the analysis of raindrop behaviours on oil slicks, which plays an important role in cleaning up oil spills.  

Despite the numerous studies spanning more than a century, the droplet impact phenomenon is still far from being fully understood \cite{Rein1993DropImpactSurfaces, Yarin2006DropImpactDynamics, C.Josserand2016DropImpactSolidSurface}. Extensive efforts and rapid progress have been made in the study of droplet impact over the last two decades \cite{Q.Deng2007ViscositySurfaceTensionBubbleEntrapment, J.M.Gordillo2010JetFormation, J.M.Gordillo2010JetBreakup, H.N.Oguz1990BubbleEntrainment, U.Jain2019VisousDrpletImpactPool, D.Liu2018DropletImpactHeatedSurface, Che2018ImmscibleImpact}. Various new techniques have been introduced to unveil the complex droplet morphologies and flow structures during the impact process, e.g., X-ray phase-contrast imaging \cite{L.V.Zhang2011DropImpactEjecta}, ultrafast interference imaging \cite{M.M.Driscoll2011InterferenceSplashing}, and brightness based laser-induced fluorescence \cite{D.B.Hann2016DrpletImpactFilmBB-LIF}. The development of the high-speed imaging technique, in particular, facilitates the detailed analysis of the impact process at a temporal resolution of sub-microseconds \cite{S.T.Thoroddsen2008DropBubble}.

During the impact of a liquid droplet onto a liquid pool, if the impact speed is large enough, the target surface is greatly deformed, and produces a hemispherical crater whose radius can be an order of magnitude larger than that of the droplet. When the crater collapses, a jet rises out of the centre of the crater, i.e., the well-known Worthington jet \cite{Rein1993DropImpactSurfaces, Worthington1897ImpactLiquidSurface}. The generation of the Worthington jet by surface waves not only occurs during the impact of a droplet onto a deep liquid pool \cite{E.Castillo-Orozco2015JetSecondaryDroplet, Zhao2011TransitionCoalescenceJet, Walker2015MiscubleJetDrop, Thoraval2016VortexBubbleEntrainment, Ray2015RegimeDropImpactPool, Pan2010DropletWetSurface, Michon2017JetDropDynamics, Manzello2002ExperimentalDropletImping, Hsiao1988WeberVortexJet}, but also in other situations such as the collapse of a bubble at a liquid-gas interface \cite{Ghabache2014BubbleBurstEjection}. A different kind of Worthington jet is induced by bubble pinch-off, which occurs in several conditions, such as the impact of a circular disk onto a liquid surface \cite{S.Gekle2009SolidImpactJetFormation}, and the impact of a droplet on a superhydrophobic solid surface \cite{D.Bartolo2006JetBubbleDropImpact}.

The fluid viscosity has a profound effect on the impact dynamics of droplets on liquid pools. Deng et al.\ studied the impact of droplets onto a liquid pool of the same fluid, and investigated the roles of viscosity and surface tension on bubble entrapment \cite{Q.Deng2007ViscositySurfaceTensionBubbleEntrapment}. They found that the capillary wave crests are weakened by viscous damping so much that the size of the entrapped bubble progressively decreases with viscosity. Marcotte et al.\ investigated the impact of ethanol droplets on pools of aqueous glycerol solutions with variable pool viscosity and air pressure \cite{F.Marocotte2018EjectaCorollaSplash}. Their results show that the liquid corolla spreading out from the impact region consists of both an ejecta and a Peregrine sheet, which, as the pool viscosity increases, evolve on separated timescales. Li et al.\ studied the intricate buckling patterns formed when a viscous drop impacts onto a miscible pool of much lower viscosity, and found that the viscous droplet will stretch into a hemispheric bowl, which would be pulled by the rotating motion around it and gradually form toroidal viscous sheets \cite{E.Q.Li2017VortexBuckingDropletImpact}. Wetting property has a profound effect on the impact dynamics of liquid droplets and solid spheres onto liquid pools. Duez et al.\  studied the solid spheres entering water at large speed \cite{Duez2007waterrepellency}, and investigated the role of wettability of the impacting sphere on the degree of splashing. They found that there is a threshold speed for air entrainment near the contact line of the impacting sphere. In their experiment, a splash was produced for the hydrophobic sphere, whereas only a tiny jet was seen during the hydrophilic sphere impacting. Jain et al.\ investigated the impacts of viscous immiscible oil droplets onto a deep pool of water \cite{U.Jain2019VisousDrpletImpactPool}. It was found that the impacting droplet created a crater, and after the retraction of the crater, there was a splash along the oil-drop rim. In this paper, we increase the viscosity ratio between the droplets and the liquid pool and tune the impact parameters in a wide range to explore the jet phenomena during droplet impact. We identify a two-jet phenomenon during the impact of a viscous droplet on a less viscous pool, i.e., two jets appear successively during one impact event.

The rest of this paper is organised as follows. The experimental details are provided in Section \ref{sec:sec2}, including the experimental setup, the fluid properties, and the high-speed photography method. The results are discussed in Section \ref{sec:sec3}, including the two-jet phenomenon and the mechanism, the evolution of the impacting droplet, and the effects of key controlling parameters. Finally, conclusions are drawn in Section \ref{sec:sec4}.

\section{Experimental section}\label{sec:sec2}

\begin{figure}[tb]
  \centering
  \includegraphics[width=0.5\textwidth]{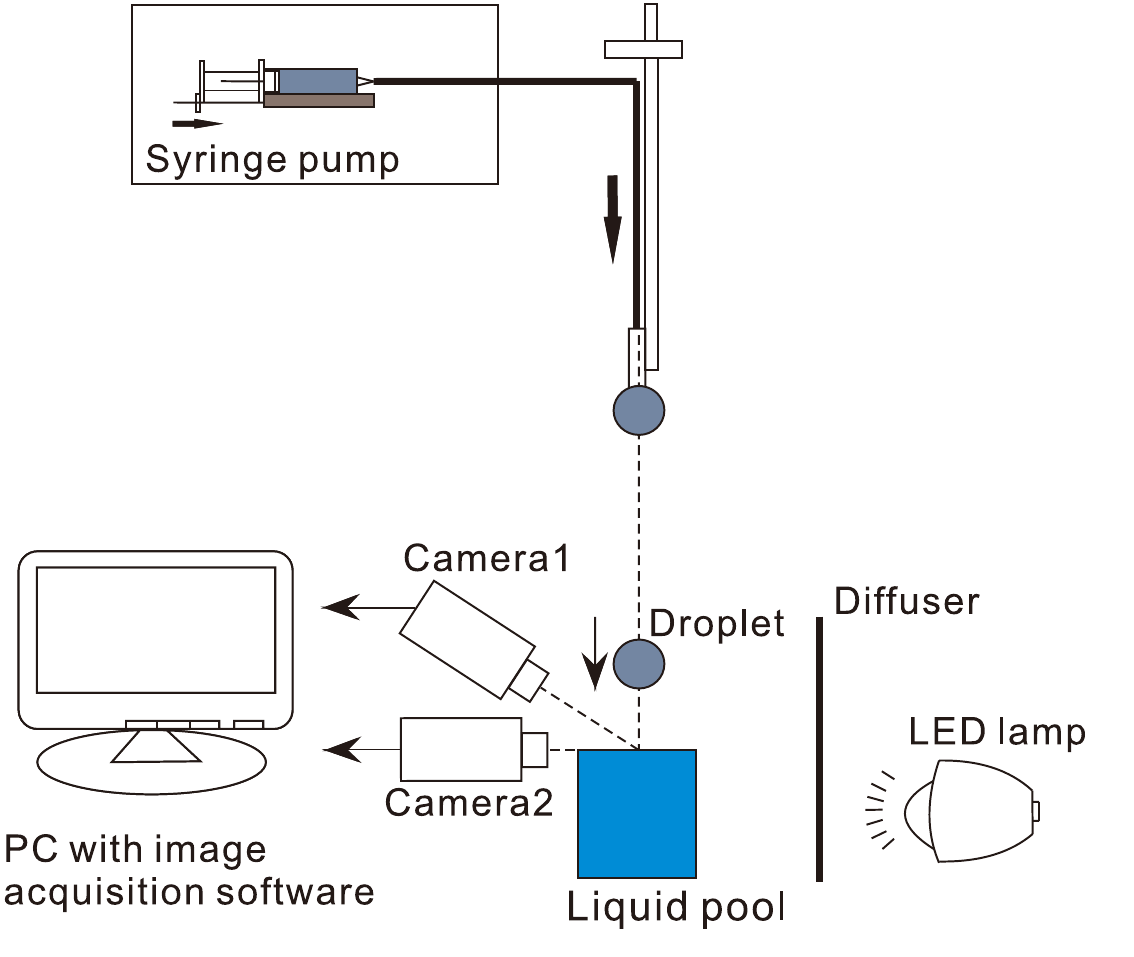}\\
  \caption{Schematic diagram of the experimental setup.}\label{fig:fig01}
\end{figure}

\begin{table}\label{tab1}
\small
\centering
\captionsetup{justification=centering}
\caption{Properties of the fluids used in this study (23 $^\circ$C).}
\begin{tabular}{lccc}
\hline
{Fluids}	&	{Density}	&{Viscosity}	&	{Surface tension}	\\
	&	$\rho$ (kg/m$^3$)	&	$\mu$ (mPa$\cdot$s)	&	$\sigma$ (mN/m)	\\
\hline
Glycerol	&	1259	&	1186	&	62.6	\\
Water+glycerol (98wt\%)	&	1255	&	784	&	62.8	\\
Water+glycerol (97wt\%)	&	1252	&	645	&	63.4	\\
Water+glycerol (96wt\%)	&	1249	&	532	&	63.4	\\
Water+glycerol (95wt\%)	&	1247	&	495	&	63.4	\\
Water+glycerol (94wt\%)	&	1244	&	374	&	63.5	\\
Water+glycerol (93wt\%)	&	1241	&	314	&	63.5	\\
Water+glycerol (92wt\%)	&	1239	&	268	&	64.0	\\
Water+glycerol (91wt\%)	&	1236	&	227	&	64.2	\\
Water+glycerol (90wt\%)	&	1233	&	194	&	64.4	\\
Water+glycerol (88wt\%)	&	1228	&	146	&	64.8	\\
Water+glycerol (85wt\%)	&	1220	&	96	&	65.4	\\
Water+glycerol (76wt\%)	&	1196	&	34	&	65.5	\\
Water+glycerol (65wt\%)	&	1166	&	14	&	66.8	\\
Water+glycerol (50wt\%)	&	1125	&	6.8	&	68.6	\\
Water	&	997	&	0.9	&	71.3	\\
\hline
    \end{tabular}
\end{table}

Water-glycerol mixtures at different glycerol concentrations were used as the working fluids, and their properties are listed in Table 1. The surface tension was measured by using a Drop Shape Analyzer (DSA100, Kruss Scientific) with the axisymmetric pendant drop method. The experimental setup to study the impact of droplets is illustrated in Fig.\ \ref{fig:fig01}. Droplets are pinched off by gravity from blunted syringe needles, then fall onto a liquid pool of lower viscosity. The size and the speed of the droplets are varied by changing the size of the needle and its height above the pool. The pool liquid is in a 40-mm-deep glass container with a 50$\times$50 mm square cross-section. The container width is more than 15 times of the droplet diameter, therefore, the effect of the container wall on the impact dynamics is negligible. After the detachment of the droplet from the syringe tip, the viscosity of the droplet rapidly dampens out any pinch-off oscillations, making the droplet be nearly spherical at impact.

The droplet shape evolution was imaged using high-speed cameras with a macro lens (Tokina 100 mm f/2.8D). Most of the high-speed images of the droplet impact processes were acquired with a high-speed camera (Photron FASTCAM SA1.1), with 5400-12000 frame per second (fps). To film the details of the surface-climbing jet, we used another high-speed camera (Photron FASTCAM SA-Z) at the speed of 25000 fps at the resolution of 1024$\times$840 pixels. Two viewing directions of the cameras were used: one was horizontal (side view), and the other was at an angle of 35$^\circ$ to the horizontal (aerial view) to obtain flow information on the interface morphology after the impact, as shown in Fig.\ \ref{fig:fig01}. An LED lamp (Hecho S5000, 60 W) combined with a diffuser was used to provide the backlight for high-speed imaging. The high-speed images were processed using a customised Matlab program to obtain the quantitative information of the impact process.

\section{Results and discussion}\label{sec:sec3}
\subsection{Two-jet phenomenon}\label{sec:sec3.1twojet}
\begin{figure*}[tb]
  \centering
  \includegraphics[width=\textwidth]{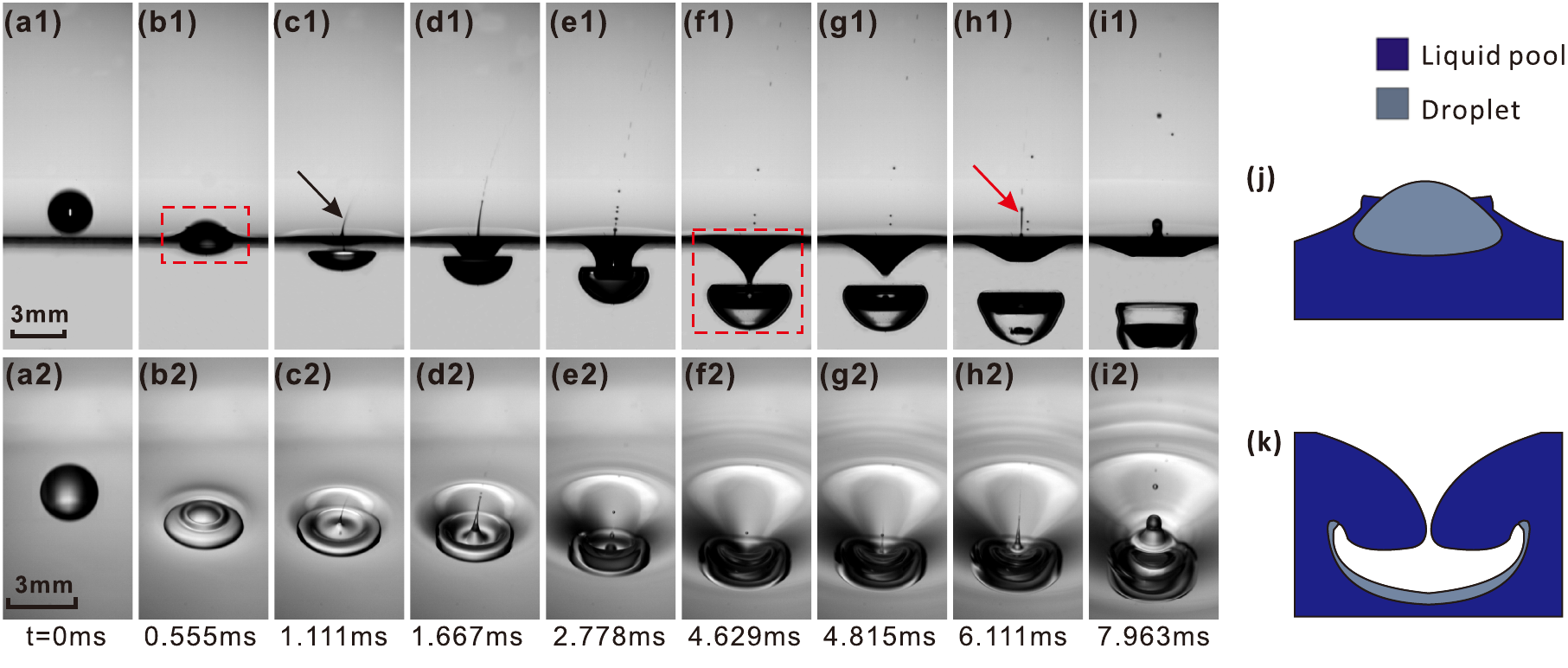}\\
  \caption{Two-jet phenomenon during one impact event. (a)-(i) Sequences of images showing the impact of a viscous droplet (glycerol/water mixture, $D=2.61$ mm, $V=3.37$ m/s, $We=429$, $\mu_d=784$ mPa$\cdot$s, $\bar{\mu}=871$) onto a water pool. The two jets are marked by black and red arrows. (a1)-(i1) Images taken from the horizontal view; (a2)-(i2) images taken from the aerial view. (j)-(k) Schematic drawings of  droplets in dashed rectangles in (b1) and (f1), respectively. See Movies 1 and 2 in Supplemental Material \cite{SMnote}.}\label{fig:fig02}
\end{figure*}

A typical impact process with the two-jet phenomenon is shown in Fig.\ \ref{fig:fig02}. The droplet fluid is a water/glycerol mixture at the glycerol concentration of 98 wt\% with a viscosity of $\mu_d=784$ mPa$\cdot$s, and the droplet diameter is $D=2.61$ mm. The liquid pool is pure water, and the droplet impact speed is $V=3.37$  m/s. When the droplet just contacts the pool surface, some pool liquid climbs along the droplet surface (see Fig.\ \ref{fig:fig02}b and the schematic drawing in Fig.\ \ref{fig:fig02}j), and rises towards the apex of the droplet. 
As the droplet continues to move downward, the bottom of the droplet becomes oblate due to the resistance of the liquid pool. Meanwhile, the liquid in the pool climbs upward along the surface of the droplet, and finally collides on the apex of the droplet, resulting in the surface-climbing jet, which is highlighted by the black arrow in Fig.\ \ref{fig:fig02}c1.

After the formation of the surface-climbing jet, the droplet continues to move downward due to its inertia, and produces a large crater behind it. Then, the crater pinches off due to the pressure of the nearby fluid (see Fig.\ \ref{fig:fig02}f and the schematic drawing in Fig.\ \ref{fig:fig02}k), producing two parts of air: the lower part forms an air bubble, while the upper part has an inverted triangular shape. Owing to the large surface tension force produced at the pinch-off point of the upper part, as shown in Fig.\ \ref{fig:fig02}g, the interface retracts immediately under the large surface tension force which accelerates the nearby fluid moving upward at a high speed. Thus, the large inertia of the fluid produces a high-speed thin liquid jet, as highlighted by the red arrow in Fig.\ \ref{fig:fig02}h1. At the same time, a large bubble is trapped inside the pool. Such liquid jet was also observed by Rein \cite{M.Rein1996TransitionCoalescingSplashing} during the impact of droplets onto a liquid pool of the same fluid. In their experiment, regular bubble entrapment is always accompanied by the ejection of a high-speed thin jet originating from the centre of the impact crater. This type of jets is termed Worthington jet induced by bubble pinch-off in this study, because it is produced by the collapse of the crater.

From the image sequences of the impact process shown in Fig.\ \ref{fig:fig02}, we can see that its difference from the impact process of the same fluid starts from the instant when the droplet contacts the pool fluid. For the droplet impact of the same fluid, the droplet deforms and the fluid at the bottom of the droplet is propelled outwards, forming a liquid lamella \cite{J.S.Lee2015VortexRingDropSplash}. In contrast, for the impact process of a viscous droplet onto a less viscous pool as shown in Fig.\ \ref{fig:fig02}, the pool fluid climbs along the droplet surface. This is because of the viscosity difference between the droplet and the pool. The time scale of surface climbing can be considered using the impact time scale ${\tau_1}$ = \emph{D}/\emph{V}, where \emph{D} is the droplet diameter and \emph{V} is the droplet impact speed. The time scale of droplet deformation can be considered from the viscous dissipation ${\tau_2} = {\rho_p{L}^{2}}/{\mu_p}$, where \emph{L} is the characteristic length, $\rho_p$ and $\mu_p$ are the density and the viscosity of the liquid pool water, respectively. It is the timescale required by viscosity to diffuse the fluid momentum over a characteristic length scale \emph{L}. To deform the droplet of diameter \emph{D} in a pool of characteristic length scale of \emph{L}, we have $\mu_p{L}$ = $\mu_d{D}$ from the balance of shear stress. Therefore, the time scale of droplet deformation in is ${\tau_2} = {\rho_p{\mu_d}^{2}{D}^{2}}/{\mu_p}^{3}$. Taking the experiment in Fig.\ \ref{fig:fig02} for example, ${\tau_2}$ is much lager than ${\tau_1}$. Therefore, it is difficult for the viscous droplet to deform in the less viscous pool in such a short instant. The surface tension force due to the large curvature upon contact only causes the pool fluid to spread along the droplet surface, instead of propelling both the pool fluid and the droplet fluid outwards to produce liquid lamella as in the same-fluid scenario.

In our experiments, after the droplet with high viscosity impacting onto the miscible pool, the surrounding liquid layer moves inward and climbs onto the droplet and forms the surface-climbing jet. However, Jain et al.\ studied the impact of oil droplets with the similar speed and viscosity onto an immiscible pool \cite{U.Jain2019VisousDrpletImpactPool} and found that there was an air cavity generated after the impact, and the liquid layer moves outward and detaches from the droplet and finally, the liquid layer developed into a splash. The difference in the impact outcomes between our result and that by Jain et al. suggests that the miscibility of droplet and pool liquid has a great influence on the moving direction of the liquid layer.

For better understanding the influence of droplet and pool liquid miscibility on the moving direction of the liquid layer and jet formation, we consider the influence of the wettability of the solid spheres on the direction of the liquid layer. Duez et al.\ studied the process of solid spheres with different static contact angle ${\theta_0}$ entering water \cite{Duez2007waterrepellency}, and they demonstrated that an air cavity was created during the impact only above a threshold speed, $V^*$. For a hydrophobic sphere (${\theta_0} \geq 90^o$), $V^*$ was found to depend on the static contact angle ${\theta_0}$ of the impact body. The threshold speed decreases with increasing the static contact angle. In an extreme case, superhydrophobic spheres produces a big splash for any impact speed. However, for a hydrophilic sphere (${\theta_0} \textless 90^o$), $V^*$ has a weak dependence on ${\theta_0}$, and they experimentally obtained $Ca_p^* = {\mu_p{V}^{*}}/{\sigma_p}$ = 0.1, where ${\mu_p}$ is the viscosity of the pool, $\sigma_p$ is the surface tension of the pool. For water pools, $V^*$ is about 7.9 m/s.

In our experiments, taking the droplet impact in Fig. 2 as an example, there is a clear droplet-pool interface and the upper half of the droplet remains spherical at the initial stage of the impact (i.e., from the instant when the droplet just touches the liquid surface to the instant when the droplet enters the pool completely). This interface indicates that the droplet and pool water do not mix significantly. Thus, the droplet can be assumed to be a soft and hydrophilic body that deforms gradually. Compared to the impact of a solid hydrophilic sphere, the upward flow of the liquid layer not only needs to overcome the gravity and the inherent viscous damping at the interface, but also converts a portion of its energy into a vortex flow inside the droplet \cite{E.Q.Li2017VortexBuckingDropletImpact}. Therefore, the viscous damping at the interface will increase and the speed threshold $V^*$ will be larger than that in the sphere impact model. For the speed considered in our experiments ($V_\text{max}=3.8$m/s, $Ca_{p,\text{max}}=0.048$), the droplet speed is much lower than the threshold. Therefore, the liquid layer follows the droplet and closes up at the pole of the droplet. As such, a surface-climbing jet is generated.

\begin{figure}[tb]
  \centering
  \includegraphics[width=0.35\textwidth]{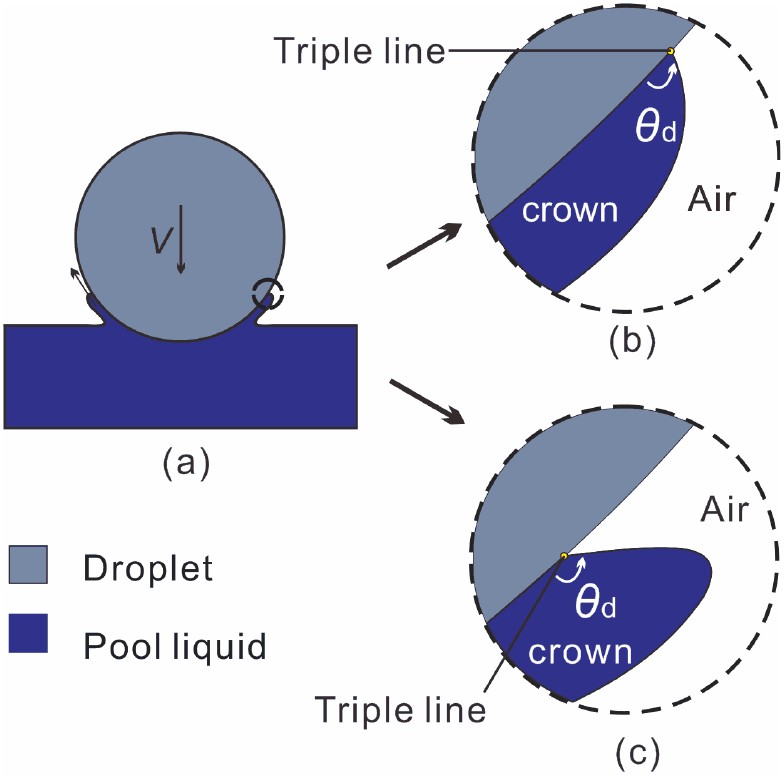}\\
  \caption{(a) Schematic diagram of droplet impact on a liquid pool. (b, c): Magnification of the triple-line region. (b) The impact speed is lower than the critical threshold speed; (c) The impact speed is higher than the threshold speed. $\theta_d$ is the dynamic contact angle, which is larger than the static contact angle $\theta_0$ for a moving triple line.}\label{fig:fig03}
\end{figure}

To compare our results with the immiscible impact studied by Jain et al.\ \cite{U.Jain2019VisousDrpletImpactPool}, we can use the contact-line stability to explain the direction of the liquid layer. The geometry is illustrated in Fig.\ \ref{fig:fig03}. In the case of the oil droplet impacting towards the water pool, the surrounding air is more likely to wrap the oil droplets than the pool water is. This situation resembles the forced wetting phenomenon, i.e., the withdrawal process of a partially wetting solid plate from one fluid to another fluid  \cite{Chan2012forcedwettingtransition}. Here the droplet surface functions as the plate and the `withdrawal' is from air into liquid pool. During the forced wetting process, it has been known that there is a threshold speed above which the triple line \cite{Sedev1991steadywettingtransition, David1999coatingonflber} (a shared boundary line of air, pool water, and droplet phase) is no longer stable, and the dynamic contact angle $\theta_d$ goes to $180^\circ$ \cite{Eggers2004Hydrodynamictheory}. Above this threshold speed, the surrounding air will start to wrap the droplet and the liquid layer will move outward as shown in Fig.\ \ref{fig:fig03}c. The threshold speed in the forced wetting process is a function of the surface wettability, and it is smaller for hydrophobic surface than for hydrophilic surfaces \cite{Duez2007waterrepellency}. Therefore, the hydrophobic oil droplet in the experiments of Jain et al.\ is above threshold, and the liquid layer is seen to detach from the droplet forming a splash. In contrast, for the same impact speed, the hydrophilic droplet in our experiment is below this threshold speed, and the liquid layer can always follow the droplet surface, as shown in Fig.\ \ref{fig:fig03}b.

\begin{figure*}[tb]
  \centering
  \includegraphics[width=0.75\textwidth]{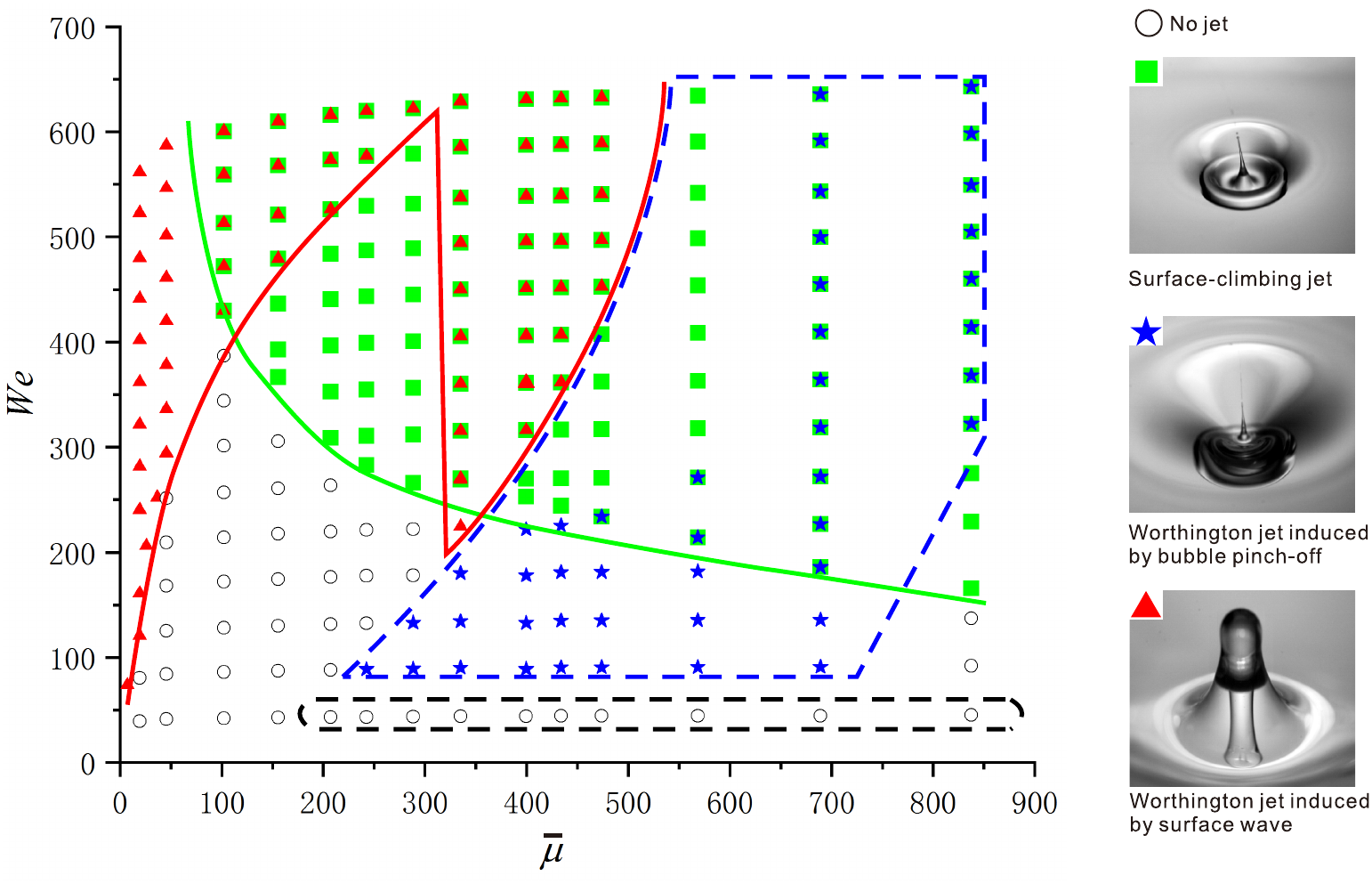}\\
  \caption{Regime map for the surface-climbing jet and the two types of Worthington jets (induced by bubble pinch-off or by surface wave). The lines in the diagram are just to guide the eyes. The green solid line and the red solid line indicate thresholds for the surface-climbing jet and for the Worthington jet, respectively. The blue dashed line indicates the region for the capturing of large bubbles, while the black dashed line for the capturing of tiny bubbles without the formation of any jet\cite{Q.Deng2007ViscositySurfaceTensionBubbleEntrapment}.
When categorising the phenomena, we regard the liquid column as a jet when there is a breakup of droplets. $\bar{\mu }\equiv {{\mu }_d}/{{\mu }_p}$, $D=2.35$ mm, and the pool liquid is pure water. }\label{fig:fig04}
\end{figure*}

\subsection{Regime map for the two-jet phenomenon in $We-\bar{\mu}$ space}\label{sec:sec3.2regimemap}
To figure out the condition in which the two-jet phenomenon occurs in one droplet impact event, a regime map of the impact outcome is produced by varying the impact speed $V$ from 0 to 3.8 m/s and the droplet viscosity from 6.8 to 784 mPa$\cdot$s, while fixing the other parameters, as shown in Fig.\ \ref{fig:fig04}. The regime map is presented in dimensionless form using the Weber number $We\equiv{\rho_d {{V}^{2}}D}/{\sigma _d}$ and the viscosity ratio between the droplet fluid and the pool fluid $\bar{\mu } \equiv \mu_d/\mu_p$, where $\rho_d$ and $\sigma_d$ are the droplet density and the droplet surface tension, respectively, $\mu_d$ and $\mu_p$ are the droplet and pool viscosities, respectively. The Weber number indicates the ratio between the inertial force and the surface tension force. As shown in Fig.\ \ref{fig:fig04}, the surface-climbing jet can be produced in a wide range of viscosity ratio. There is a threshold of $We$ for the formation of the surface-climbing jet (as highlighted by the green solid line), and as the viscosity ratio increases, the threshold of $We$ decreases. More details about the dependence of the surface-climbing jet on the droplet $We$ and the viscosity ratio $\bar{\mu}$ will be discussed later in Subsections \ref{sec:sec3.3dropletSpeed} and \ref{sec:sec3.4dropletViscosity}.

In the regime map shown in Fig.\ \ref{fig:fig04}, both the red triangles and blue stars represent regions of the Worthington jet. The difference is that the jet produced in the red triangle region is caused by free-surface waves, while the jet produced in the blue star region is caused by bubble pinch-off. Since both of them are produced by the collapse of the crater,  both are termed the Worthington jet. As for the Worthington jet induced by bubble pinch-off, the pinch-off process produces a singularity point and therefore, the jet is much thinner with a higher speed than the Worthington jet induced by surface wave (see the example images in Fig.\ \ref{fig:fig04}). It should be noted that the capture of a bubble does not always lead to the formation of a Worthington jet  (see Movie 3 in Supplemental Material \cite{SMnote}). The region of bubble capturing is indicated by the blue dashed line in Fig.\ \ref{fig:fig04}. In the upper-left part of this region of bubble capture, the cone angle of the inverted triangular crater is too large, so there is not enough energy for the crater to retract to form a Worthington jet.

The formation of the Worthington jet requires a certain amount of droplet kinetic energy, i.e., a threshold of $We$, as highlighted by the red solid line in Fig.\ \ref{fig:fig04}. At low viscosity ratios ($\bar{\mu }<330$), the threshold of $We$ increases as $\bar{\mu }$ increases. When the viscosity ratio $\bar{\mu }$ increases beyond 330, the threshold of $We$ suddenly decreases, causing the $We$ range of generating the Worthington jet to expand. This is closely related to the aforementioned bubble capture phenomenon at $\bar{\mu }>330$. Within this viscosity range, the droplet collides with the liquid pool to generate the surface-climbing jet and continues to sink, meanwhile, the liquid at the edge of the crater converges inward. Under such a viscosity condition, when the droplet speed is low, the surrounding liquid above the crater converges and traps a bubble (see Movie 3 in Supplemental Material \cite{SMnote}). However, when the droplet speed is high enough, the radius of the crater increases and the liquid could not completely close the crater so that it can provide more energy to retract, resulting in a stronger surface-wave Worthington jet (see Movie 4 in Supplemental Material \cite{SMnote}).

Because the Worthington jet has been studied extensively in the literature, next, we mainly focus on the surface-climbing jet. We study the effects of the key parameters controlling the impact process, such as the droplet impact speed, the droplet viscosity, the droplet size, and the surface tension of the liquid pool.

\subsection{Effect of droplet speed}\label{sec:sec3.3dropletSpeed}
\begin{figure}[tb]
  \centering
  \includegraphics[width=\columnwidth]{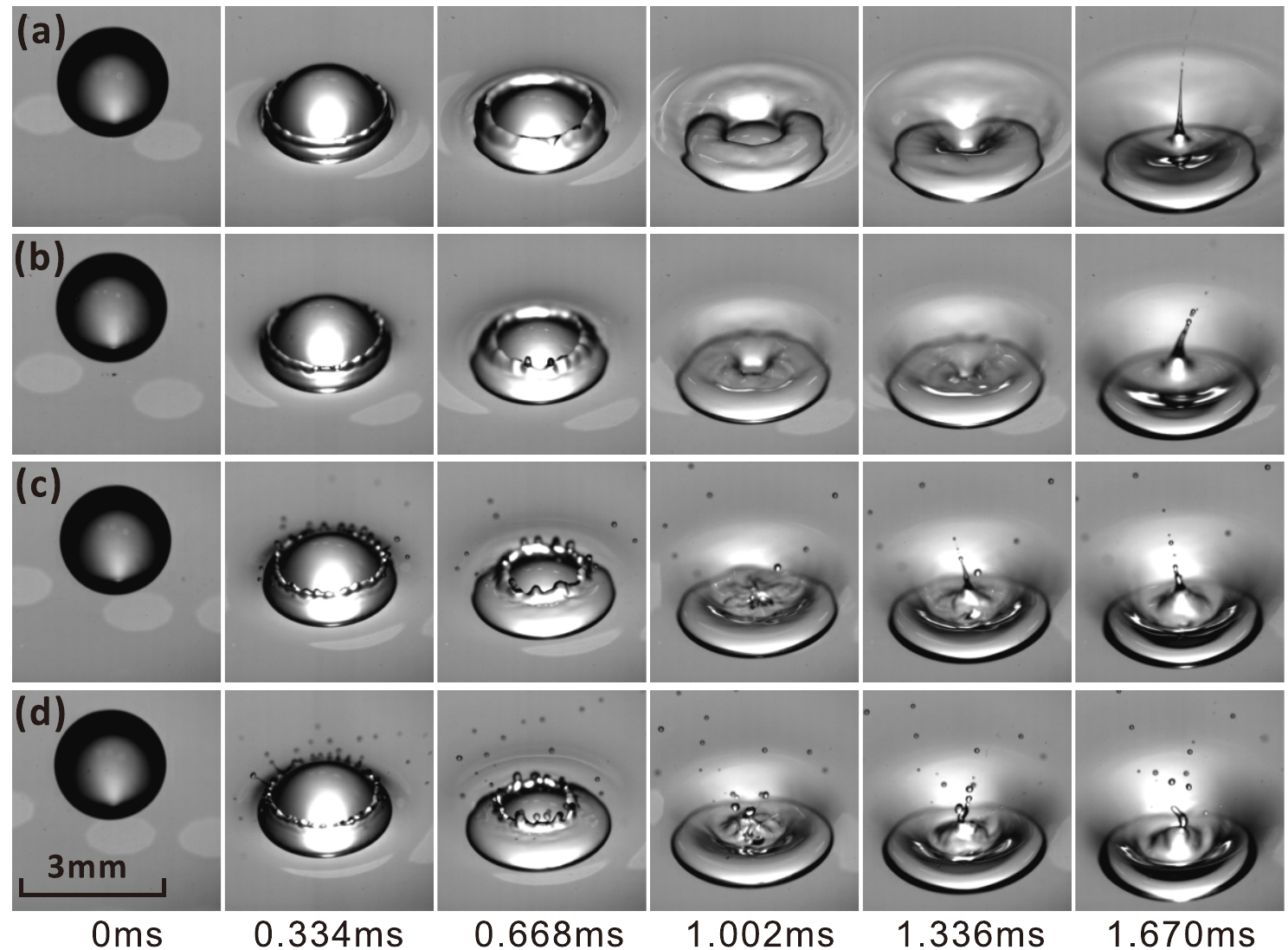}\\
  \caption{Sequences of images showing the effect of increasing the impact speed for a glycerol droplet impacting a pool of water, $D=2.35$ mm, $\mu_d=1186$ mPa$\cdot$s, $\bar{\mu}=1318$.  (a) $V=2.33$ m/s, $We=257$; (b) $V=2.62$ m/s, $We=324$; (c) $V=2.91$ m/s, $We=400$; and (d) $V=3.20$ m/s, $We=484$. }\label{fig:fig05}
\end{figure}

Figure \ref{fig:fig05} shows image sequences for the effect of varying the impact speed. It is found that changing the impact speed of the droplet will affect the speed, the morphology and the formation time of the surface-climbing jet. When the impact speed is small, only the Worthington jet induced by surface wave is produced, as shown in Fig.\ \ref{fig:fig05}a. A liquid layer is generated around the droplet and climbs along the droplet surface, but finally fails to reach the apex of the droplet to form a surface-climbing jet. If the droplet impact speed increases, the liquid layer can climb higher, as shown in Fig.\ \ref{fig:fig05}b. As the impact speed increases further, the liquid layer finally converges at the apex of the droplet and produces a surface-climbing jet (see Fig.\ \ref{fig:fig05}c). Increasing the impact speed also advances the formation time of the surface-climbing jet. At 1.336 ms, a distinct jet had formed for $V=2.91$ m/s as shown in Fig.\ \ref{fig:fig05}c, but for $V=2.62$ m/s in Fig.\ \ref{fig:fig05}b, the liquid layer around the droplet just reached the apex of the droplet. When the impact speed increases further as shown in Fig.\ \ref{fig:fig05}d  (also see Movie 5 in Supplemental Material \cite{SMnote}), the liquid layer becomes undulated and then many secondary droplets are generated from the rim of the liquid layer. As a consequence, the undulated liquid layer inhibits the formation of the surface-climbing jet.

\begin{figure}[tb]
  \centering
  \includegraphics[width=0.45\textwidth]{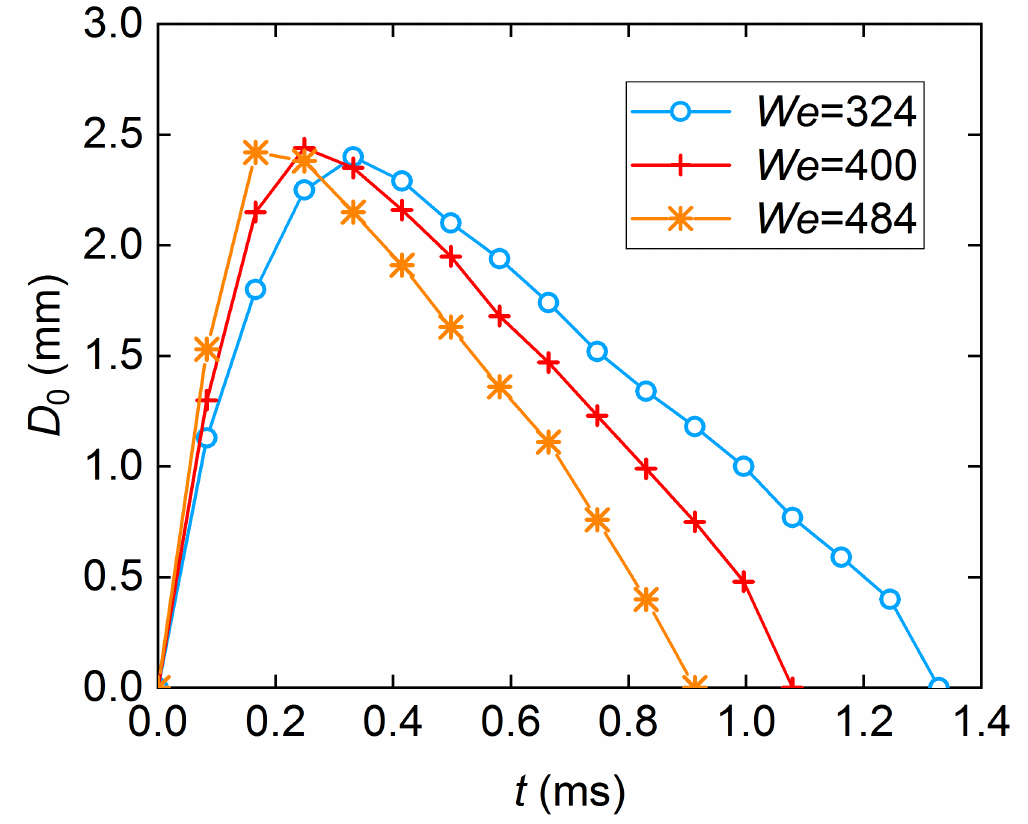}\\
  \caption{Diameter of the rim of the liquid layer ($D_0$) obtained from high-speed images corresponding to Fig.\ \ref{fig:fig05}.}\label{fig:fig06}
\end{figure}

We obtained the diameter evolution of the rim of the liquid layer (in Fig.\ \ref{fig:fig05}) through image processing as shown in Fig.\ \ref{fig:fig06}. From this plot, we can more intuitively see the effect of the impact speed on the jet formation. After the impact, the yellow curve with the highest \emph{We} (\emph{We} =484, corresponding to Fig.\ \ref{fig:fig05}d) is ahead of the other two curves. It first reaches a maximum value ($D_\text{0max}$), and then decreases to 0, indicating the closure of the liquid layer at the droplet apex and the formation of the surface-climbing jet. In contrast, the curves with lower \emph{We} are slightly delayed. In addition, $D_\text{0max}$ is slightly larger than the droplet diameter before the impact ($D=2.35$ mm) in all the cases, because the droplet is deformed upon the impact and slightly spreads in the horizontal direction.

\subsection{Effect of droplet viscosity}\label{sec:sec3.4dropletViscosity}

\begin{figure}[tb]
  \centering
  \includegraphics[width=\columnwidth]{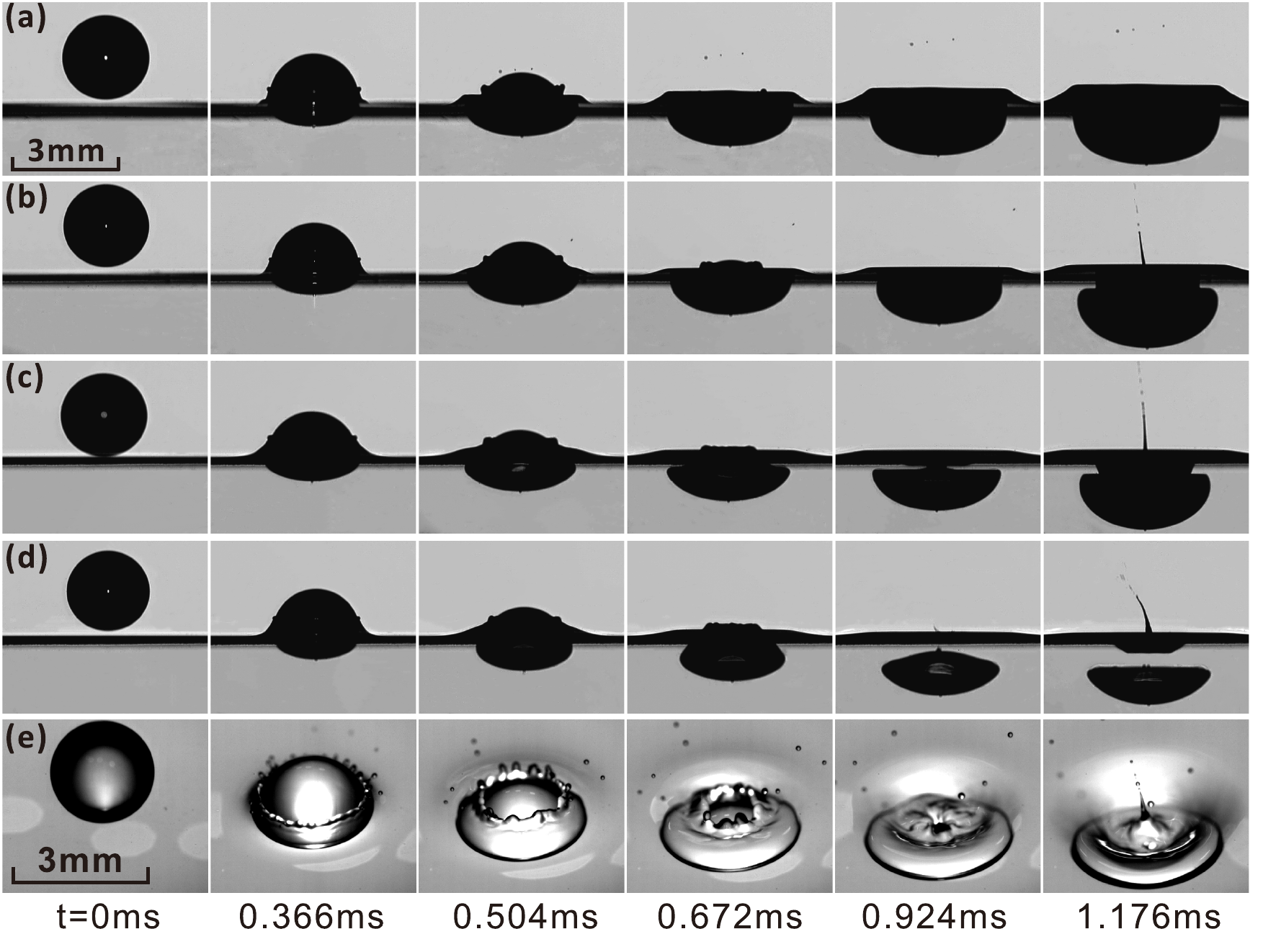}\\
  \caption{Sequences of images showing the effect of increasing the viscosity of the droplet, for droplets of various viscosities impacting a pool of water, $D=2.35$ mm, $V=3.13$ m/s, $We=429\sim463$. 
  (a) $\mu_d=96$ mPa$\cdot$s, $\bar{\mu}=107$; (b) $\mu_d =314$ mPa$\cdot$s, $\bar{\mu}=349$; (c) $\mu_d =645$ mPa$\cdot$s, $\bar{\mu}=717$; (d-e) $\mu_d =1186$ mPa$\cdot$s, $\bar{\mu}=1318$. (a)-(d) Images taken from the horizontal view; (e) images taken from the aerial view.}\label{fig:fig07}
\end{figure}

\begin{figure}[tb]
  \centering
  \includegraphics[width=0.45\textwidth]{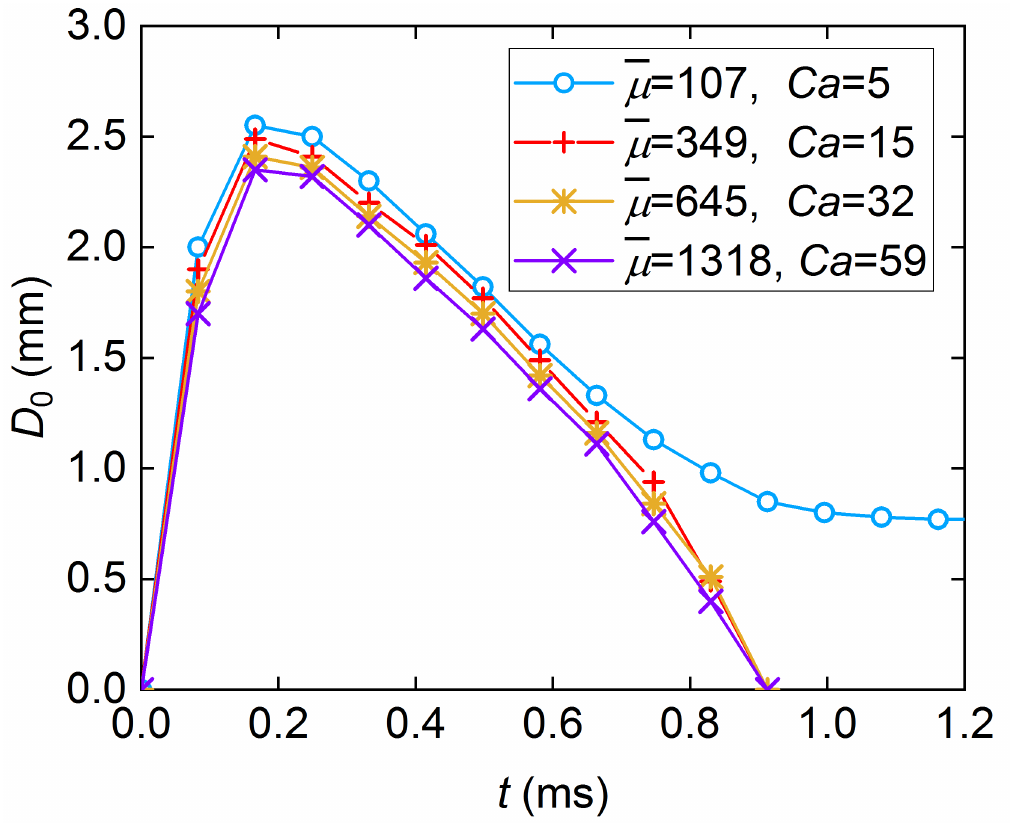}\\
  \caption{Diameter of the rim of the liquid layer obtained from high-speed images corresponding to Fig.\ \ref{fig:fig07}.}\label{fig:fig08}
\end{figure}

The importance of the droplet viscosity can be seen by comparing viscous force and surface tension force, i.e., the Capillary number \emph{Ca}. In our experiments, the Capillary number changes from $Ca \sim O(10^{-1})$ for droplets of water-glycerol mixture (50 wt$\%$) to $Ca \sim O(10)$ for droplet of pure glycerol. The effect of the droplet viscosity on the surface-climbing jet is shown in Fig.\ \ref{fig:fig07} ($Ca=5 \sim 59$). It can be found that the change in the droplet viscosity affects the deformation of the droplet, which consequently affects the formation and morphology of the surface-climbing jet. From Fig.\ \ref{fig:fig07}a, we can see that the liquid layer is produced around the droplet when the droplet viscosity is small, and moves inwards to the center of the impact point, but this liquid layer fails to form a surface-climbing jet. We can see it more clearly from the blue curve in Fig.\ \ref{fig:fig08} ($\bar{\mu}=107$). The curve tends to be flat around 0.9 ms after the impact, indicating that the rim diameter of the liquid layer remains almost unchanged after that moment. Because of the low viscosity of the droplet, the droplet deforms greatly and spread along the surface of the pool liquid to form a toroidal viscous sheet \cite{E.Q.Li2017VortexBuckingDropletImpact}. The droplet starts to mix with the pool liquid and there is no clear interface between them. Under this condition, the droplet can no longer be assumed to be a slightly deformed hydrophilic solid body, and there is no jet generated. As the viscosity of droplet increases, a surface-climbing jet gradually forms at the apex of the droplet. This is because the viscosity of the droplet damps the deformation of the droplet during the impact. Before the formation of the surface-climbing jet, the droplet can maintain a flat spherical shape as shown in Fig.\ \ref{fig:fig07}e. After the jet formation, the droplet only deforms into a hemispherical bowl-like structure and there is a clear interface between the droplet and the surrounding liquid. Therefore, the hydrophilic solid body model is still suitable for droplets impacting deep pools at this viscosity. It can also be seen in Fig.\ \ref{fig:fig08} that when the \emph{Ca} number is greater than 15, the diameter $D_0$ decreases to 0, indicating the closure of the liquid layer and the formation of the surface-climbing jet. The larger the \emph{Ca} number is, the smaller the $D_\text{0max}$ will be. This is because when the viscosity is increased, the droplet becomes more similar to a solid sphere, and the droplet deformation is smaller in the horizontal direction. In addition, the viscosity of the droplet also has a great influence on the liquid layer. Although the inward motion of the liquid layer does not change as \emph{Ca} increased, when \emph{Ca} exceeds the splash threshold \cite{Yarin1995splashthreshold}, the rim of the liquid layer becomes nonuniform because the surface tension and inertial effects associated with the liquid layer destabilize the rim. The transverse instability in the rim of the liquid layer \cite{Roisman2006sprayimpact} leads to the formation of cusps and filaments \cite{Roisman2007breakupcrown} which finally breaks up \cite{Jens1997Nolinear, Maromttant2004Fragmentation} and leads to the formation of many secondary droplets, as shown in Fig.\ \ref{fig:fig07}e. The loss of integrity of the climbing liquid layer has a certain inhibitory effect on the jet. Therefore, the jet becomes curved (see Fig.\ \ref{fig:fig07}d) and the jet speed is reduced.

\subsection{Effect of surface tension}\label{sec:sec3.5surfaceTension}
Surface tension plays an important role in droplet dynamics, and it can even produce a jet under the Marangoni stress when a droplet coalesces with a fluid of lower surface tension \cite{F.BlanchetteCoalescenceDropInterface, K.Sun2018MarangoniFlowCoalescence}. It should be noted that the formation of the surface-climbing jet is completely different from the jet formation due to the Marangoni stress. Even though there is a slight difference in the surface tension between the droplet fluid and the pool fluid (see Table 1), the Marangoni stress is much smaller than the inertia stress and the viscous stress. This can be estimated from the Weber number and the Capillary number, which are respectively $We \equiv {\rho_d {{V}^{2}}D}/{\sigma _d }=592$, $Ca \equiv {\mu_d V}/{\sigma_d }=42$ for the case shown in Fig.\ \ref{fig:fig02}. This indicates the smaller effect of the surface tension force relative to the inertia and viscous forces, and hence the small effect of the Marangoni stress. In addition, the surface tension of the droplet fluid is smaller than the surface tension of the pool fluid in this study, while the jet induced by the Marangoni stress relies on the fact that the droplet has a larger surface tension than the pool \cite{K.Sun2018MarangoniFlowCoalescence}.

\begin{figure}[tb]
  \centering
  \includegraphics[width=\columnwidth]{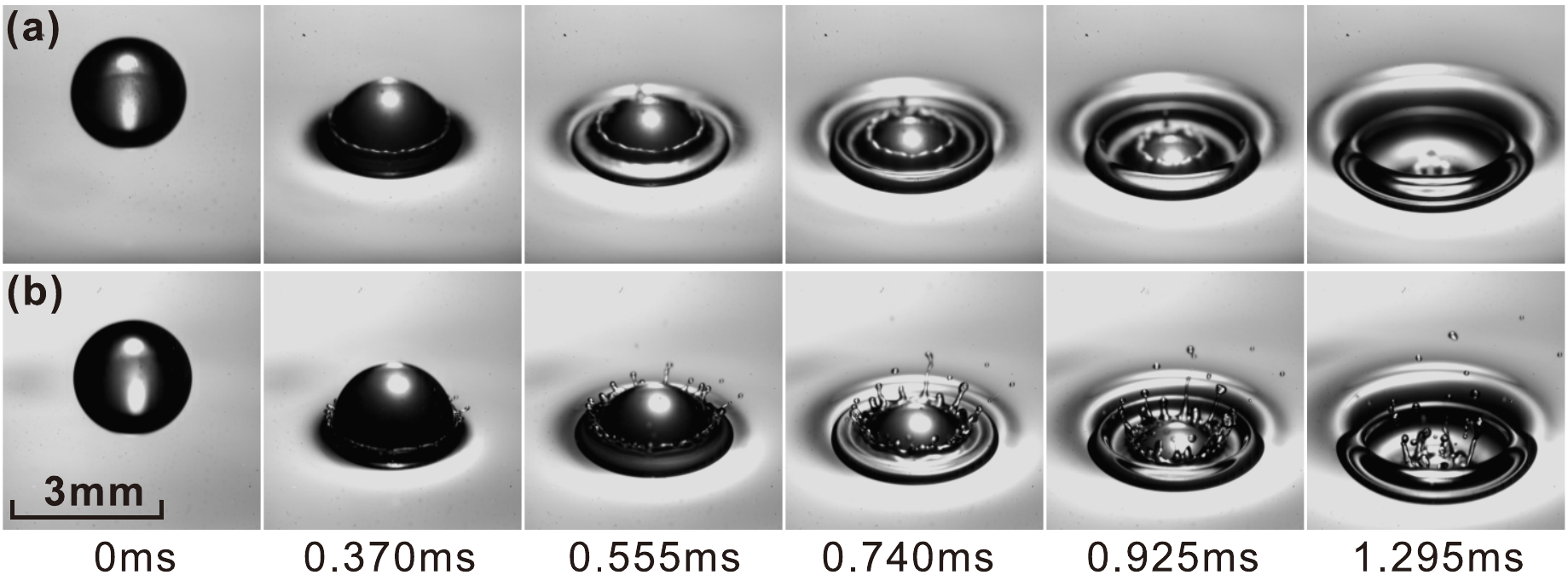}\\
  \caption{Sequences of images showing the effect of changing the surface tension of the pool fluid. $D=2.35$ mm, $V=2.62$ m/s, $We=309$, $\mu_d=194$ mPa$\cdot$s, $\bar{\mu}=216$. (a) The liquid pool is pure water, $\sigma _p=71.3$ mN/m. (b) Surfactant SDS is added to the pool at a concentration of 2 wt\%, $\sigma _p=44.2$ mN/m.}
  \label{fig:fig09}
\end{figure}

To further study the effect of the surface tension on the surface-climbing jet, we changed the surface tension of the pool fluid while fixing other parameters unchanged. This is achieved by adding a surfactant (Sodium dodecyl sulfate, SDS) into the pool (water) at a concentration of 2 wt\%, which reduces the surface tension from 71.3 to 44.2 mN/m. It can be seen from Fig.\ \ref{fig:fig09} that as the surface tension of the pool fluid decreases, the liquid layer that climbs along the droplet surface becomes more undulated and there are many secondary droplets generated from the rim of the liquid layer. This is because the decreasing of the surface tension enhances the rim instability \cite{Roisman2006sprayimpact}. Finally, the loss of integrity of the liquid layer has a certain inhibitory effect on the jet. However, it still fails to climb to the droplet apex and produces the surface-climbing jet. That is to say, reducing the surface tension of the pool liquid could not promote the formation of a surface-climbing jet.

\subsection{Effect of droplet size}\label{sec:sec3.6dropletSize}
\begin{figure}[tb]
  \centering
  \includegraphics[width=\columnwidth]{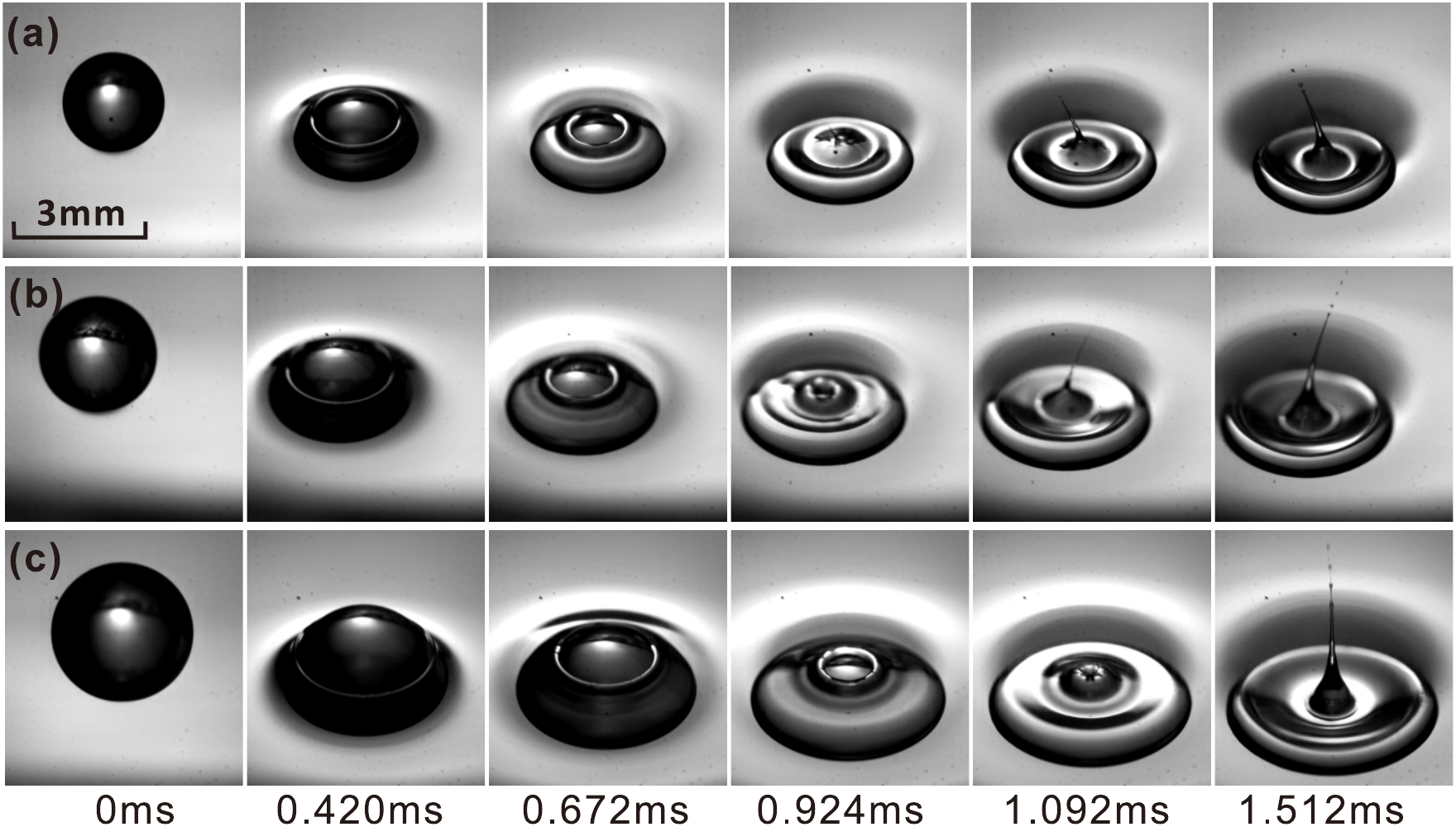}\\
  \caption{Sequences of images showing the effect of changing droplet size, for droplets impacting on a pool of pure water, $V=2.62$ m/s, $\mu_d=1186$ mPa$\cdot$s, $\bar{\mu}=1318$. (a) $D=2.32$ mm, $We=320$; (b) $D=2.70$ mm, $We=373$; (c) $D=3.24$ mm, $We=447$.}\label{fig:fig10}
\end{figure}

\begin{figure}[tb]
  \centering
  \includegraphics[width=0.4\textwidth]{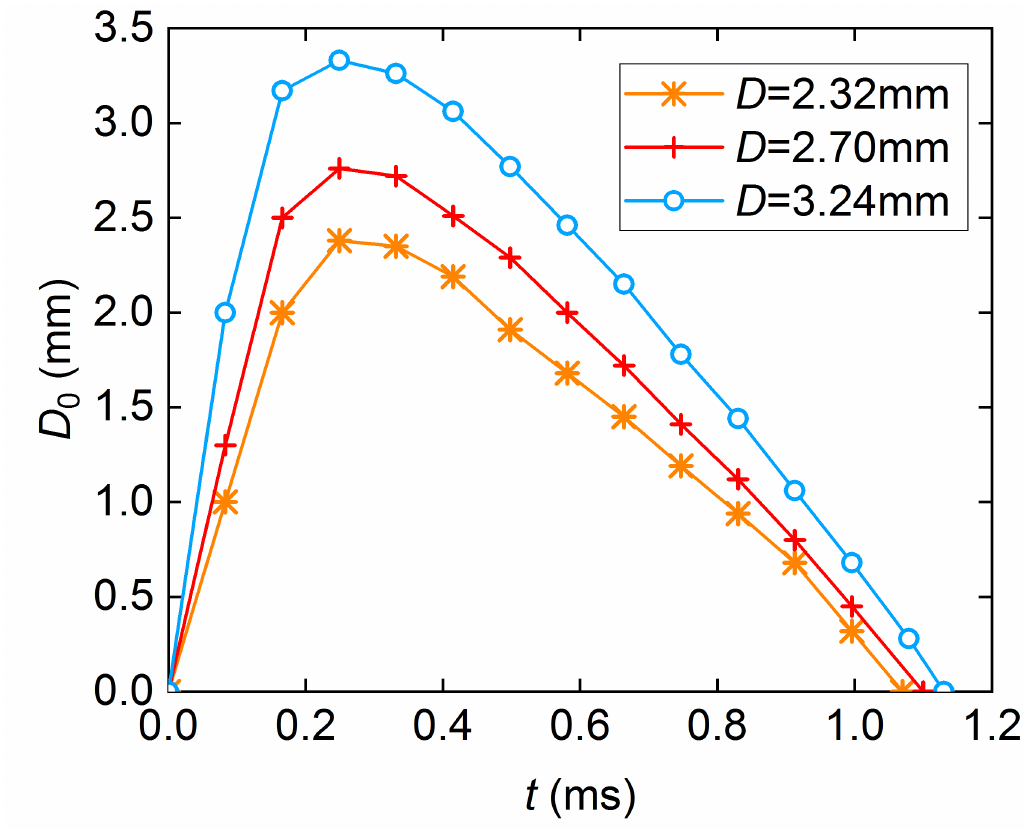}\\
  \caption{Diameter of the rim of the liquid layer obtained from high-speed images corresponding to Fig.\ \ref{fig:fig10}.}\label{fig:fig11}
\end{figure}

To study the effect of the droplet size on the surface-climbing jet, we changed the droplet size by using different syringe needles, while fixing other parameters unchanged. The droplet diameter is varied from 2.32 to 3.24 mm, and the effect on the surface-climbing jet is shown in Fig.\ \ref{fig:fig10}. It is found that changing the droplet size affects the speed, the direction, and the formation time of the surface-climbing jet. Increasing the droplet size will increase the speed of the surface-climbing jet. When droplet size $D$ is increased from 2.32 to 3.24 mm, the speed of the surface-climbing jet $V_\text{jet}$ (measured from high-speed images at the instant of jet formation) increases from 10.2 to 19.6 m/s. Increasing the droplet size $D$ also changes the direction of the surface-climbing jet. The jet is ejected at a certain angle of inclination when the droplet size is small (see Figs.\ \ref{fig:fig10}a-b). However, when the droplet size is as long as $D=3.24$ mm, the jet can keep vertical in the process (see Fig.\ \ref{fig:fig10}c, also see Movie 6 in Supplemental Material \cite{SMnote}). Increasing the droplet size also delays the formation of the surface-climbing jet. At 1.092 ms, jets had formed for $D =$2.32 and 2.70 mm (as shown in Figs.\ \ref{fig:fig10}a-b respectively), but for $D=3.24$ mm in Fig.\ \ref{fig:fig10}c, the liquid layer around the droplet just reached the apex of the droplet. We can see the evolution of the liquid layer more clearly from its rim diameter in  Fig.\ \ref{fig:fig11}. In the three cases, the maximum diameter of the rim increases with the droplet size and the rims all reach $D_\text{0max}$ at about 2.3 ms after impact. However, the yellow curve for the smallest droplet size ($D = 2.32$ mm) first decreases to 0, indicating the first formation of the surface-climbing jet. Increasing the droplet size means increasing the kinetic energy of the droplet, and the toroidal vortex on the side of the droplet is strengthened, so the speed of the surface-climbing jet increases. Meanwhile, the distance over which the liquid layer climbs is also increased with the droplet size. As a result, the formation of the surface-climbing jet is delayed.

\subsection{Speed of the surface-climbing jet}\label{sec:sec3.7jetSpeed}
\begin{figure}[tb]
  \centering
  \includegraphics[width=0.42\textwidth]{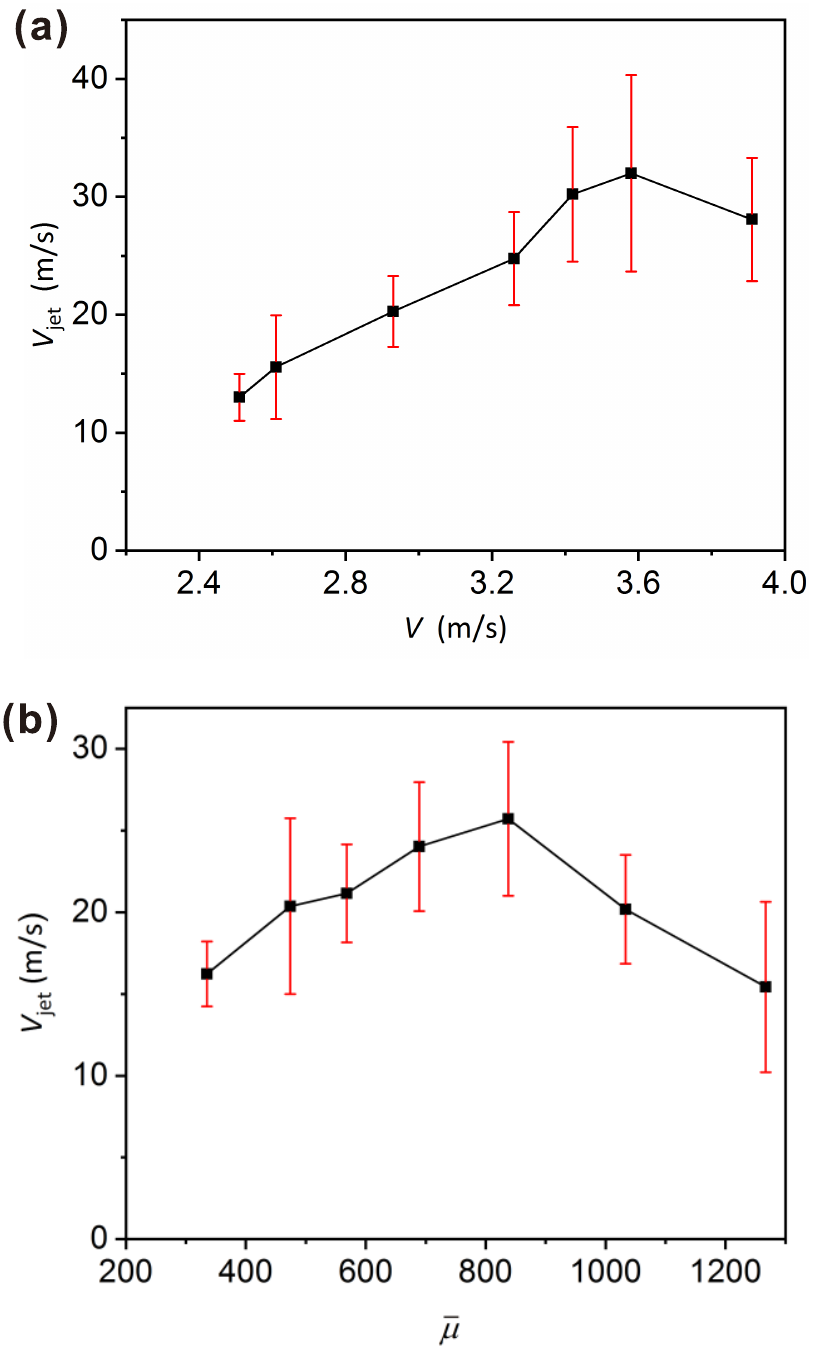}\\
  \caption{(a) Relationship between the speed of the surface-climbing jet ($V_\text{jet}$) and the impact speed of droplets ($V$).  $\mu_d =314$ mPa$\cdot$s;  (b) relationship between the speed of the surface-climbing jet ($V_\text{jet}$) and the droplet-pool viscosity ratio ($\bar{\mu }$). $V=2.62$ m/s.  For each point in the plots, the experiment was repeated 10 times, and the error bars show the standard deviations. }\label{fig:fig12a}
\end{figure}

To further characterise the surface-climbing jet produced in the impact process, we measured the speed of the surface-climbing jet from the high-speed images. The variation of the speed of the surface-climbing jet with the impact speed is shown in Fig.\ \ref{fig:fig12a}a. We can see that the surface-climbing jet speed $V_\text{jet}$ is an order of magnitude higher than the impact speed $V$. The speed of surface-climbing jet has a tendency to increase first and then decrease as the impact speed increases. This is because increasing the droplet impact speed means the inertia force of the droplet becomes larger, and the flow velocity around the droplet is increased so that the liquid layer has more energy to converge at the droplet apex, resulting in the increase in the jet speed. However, when the impact speed continues to increase (see Fig.\ \ref{fig:fig10}d), the liquid layer becomes nonuniform because of the capillary force. The loss of integrity of the liquid layer inhibits the formation of the surface-climbing jet, thus the jet speed reduces. This also verifies our aforementioned results in Sections \ref{sec:sec3.3dropletSpeed} and \ref{sec:sec3.4dropletViscosity} that the undulated liquid layer suppresses the surface-climbing jet at high impact speeds.

The variation of the jet speed with the viscosity ratio is shown in Fig.\ \ref{fig:fig12a}b. It can be found that at small viscosity ratios ($\bar{\mu }<830$), the jet speed $V_\text{jet}$ increases with $\bar{\mu }$. However, when $\bar{\mu }$ becomes larger than 830, $V_\text{jet}$  starts to decrease. This is also caused by the undulation of the liquid layer climbing along the droplet surface (see the image sequence in Fig.\ \ref{fig:fig10}e).

\section{Conclusions}\label{sec:sec4}
The impact of viscous droplets onto a less-viscous liquid pool is studied experimentally, and it is found that two jets can occur successively during one impact event. The first jet, named surface-climbing jet, is generated from the liquid layer which is induced by the vortex ring around the droplet climbing along the droplet surface and converging on the droplet apex. The second jet is the traditional Worthington jet. According to whether a bubble is trapped in the liquid pool, the second jet after the retraction of the crater can be induced either by surface wave or by bubble pinch-off. The pinch-off of a bubble produces a singularity point, and generates a high-speed Worthington jet. The speed of the surface-climbing jet is measured from the high-speed images, and it is found to be one order of magnitude higher than the impact speed. The effects of the impact speed, the droplet viscosity, the droplet size, the surface tension of the liquid pool, and the droplet-pool miscibility on the surface-climbing jet are analysed. 

This study focuses mainly on jet formation during the impact of viscous droplets onto a less-viscous liquid pool. There are many open questions in this field, such as the detailed measurement of the liquid layer climbing along the droplet surface, the analysis of the transition from the surface-climbing liquid layer to the jet, and the full numerical simulation of the impact process. A clear understanding of the impact process may not only provide physical insights into the mechanism of droplet/jet dynamics, but also be helpful in the optimisation of this process in relevant applications.

\begin{acknowledgments}
This work is supported by the National Natural Science Foundation of China (Grant No.\ 51676137), the Natural Science Foundation of Tianjin City (Grant No.\ 16JCYBJC41100), and the National Natural Science Funds for Distinguished Young Scholar (No.\ 51525603).
\end{acknowledgments}

\bibliography{TwoJetDropletImpact}
\end{document}